\documentstyle[graphicx,aps,prl,twocolumn]{revtex}

\newcommand{\myscalebox}[1]{\scalebox{0.4}[0.45]{#1}}
\newcommand{\captionA}
{Stiffness energy$\Delta$ as function of system size $L$. The line
represents the function $\Delta(L)=aL^{\Theta_S}$ with $\Theta_S=0.19(2)$. 
The inset shows the same figure on log-log scale. The increase of $\Delta$
with system size indicates, that in 3d Ising spin glasses an
ordered phase exists below a non-zero temperature $T_c$.
}

\newcommand{\captionB}
{Average energy as function of simulation parameters $M_i*\nu*n_{\min}$ 
for 90 systems of size $L=6$.
With increasing numerical effort the energy lowers. For the two
rightmost points ($M_i=64,128$; $\nu=4$; $n_{\min}=2$) no further decrease
of the energy is possible. A comparison with exact ground state calculations
using a
 Branch-and-Cut algorithm confirms that in fact for all 90 systems the true
ground states are found!
}

\newcommand{\captionC}
{Average stiffness energy $\Delta$ as function of simulation parameters
$M_i*\nu*n_{\min}$ for 90 systems of size $L=6$. This quantity is
more sensitive to changes in the simulation parameters than the energy.
}

\newcommand{\captionD}
{Average stiffness energy $\Delta$
as function of sample size $N_L$ for three different
sets of simulation parameters for system size $L=6$. If the calculated
states are close to the true ground state (set (16,4,2)) the resulting 
stiffness energy is very close to the true value and converges towards it
with increasing sample size. (The 90 samples from Fig. \ref{fig_test_delta}
are not included in the samples here, so the values for $N_L=100$ are
different from the results above.)
}

\newcommand{\figA}{
\begin{figure}[ht]
\begin{center}
\myscalebox{\includegraphics{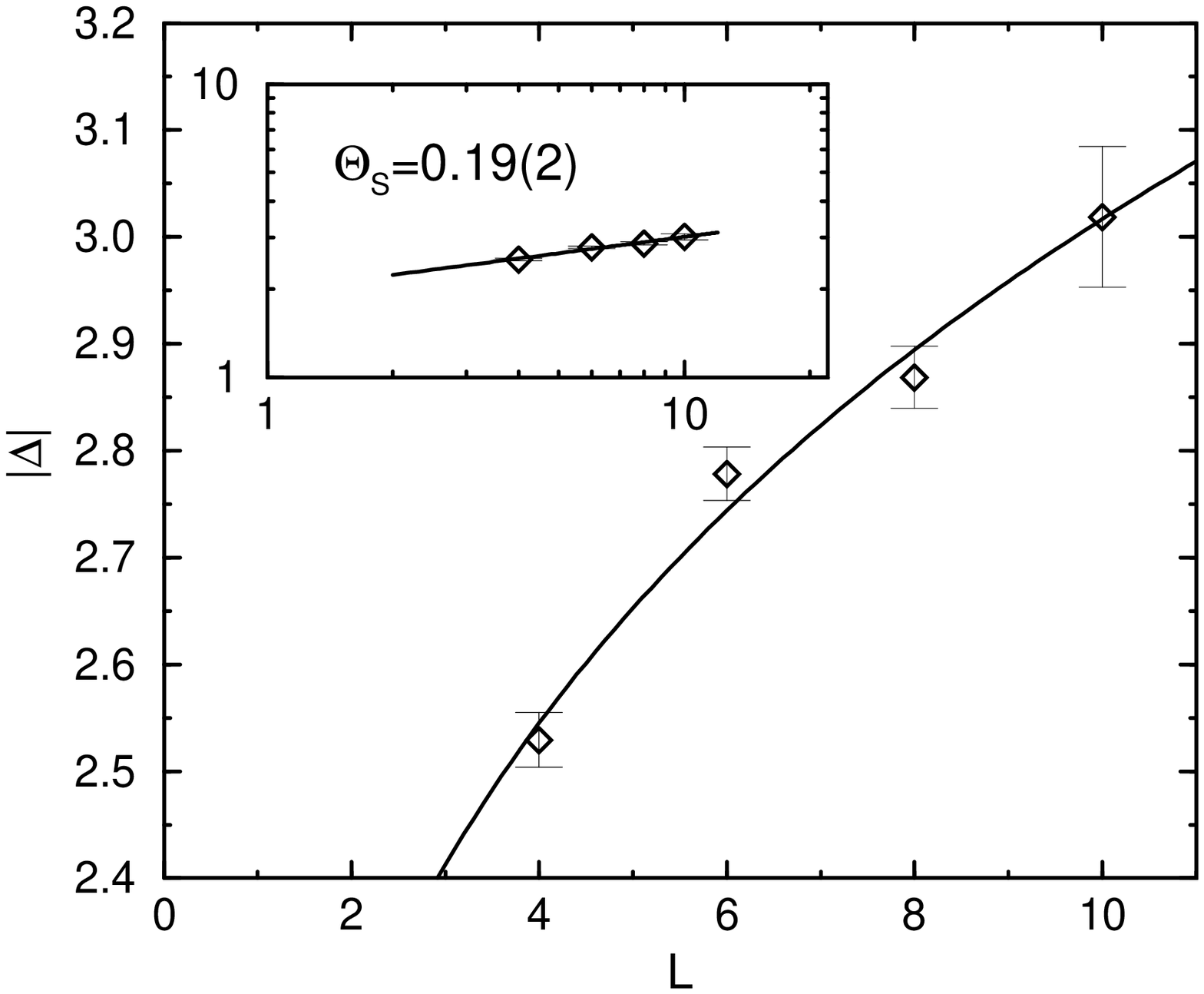}}
\end{center}
\caption{\captionA}
\label{fig_stiff_L}
\end{figure}
}

\newcommand{\figB}{
\begin{figure}[ht]
\begin{center}
\myscalebox{\includegraphics{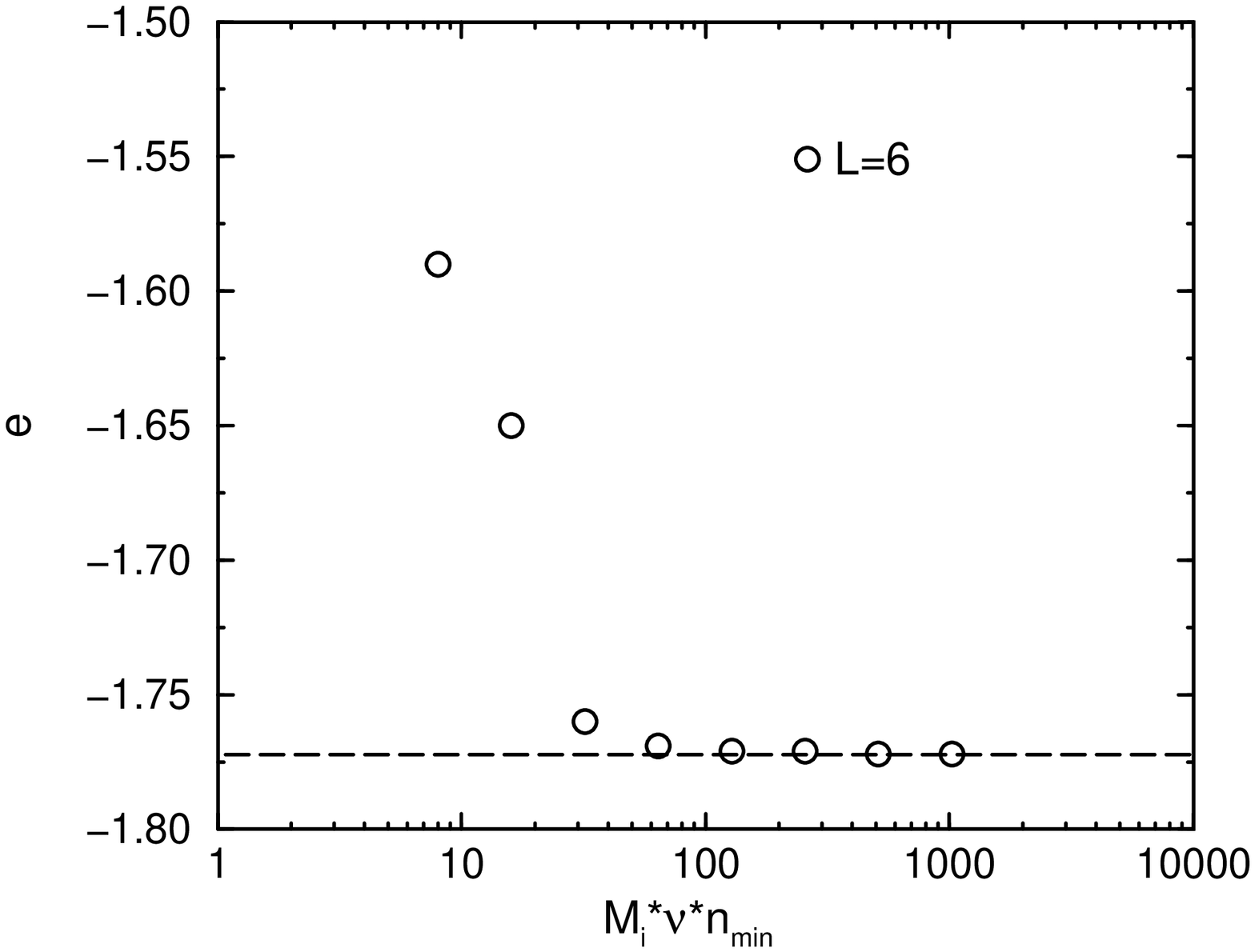}}
\end{center}
\caption{\captionB}
\label{fig_test_e}
\end{figure}
}

\newcommand{\figC}{
\begin{figure}[ht]
\begin{center}
\myscalebox{\includegraphics{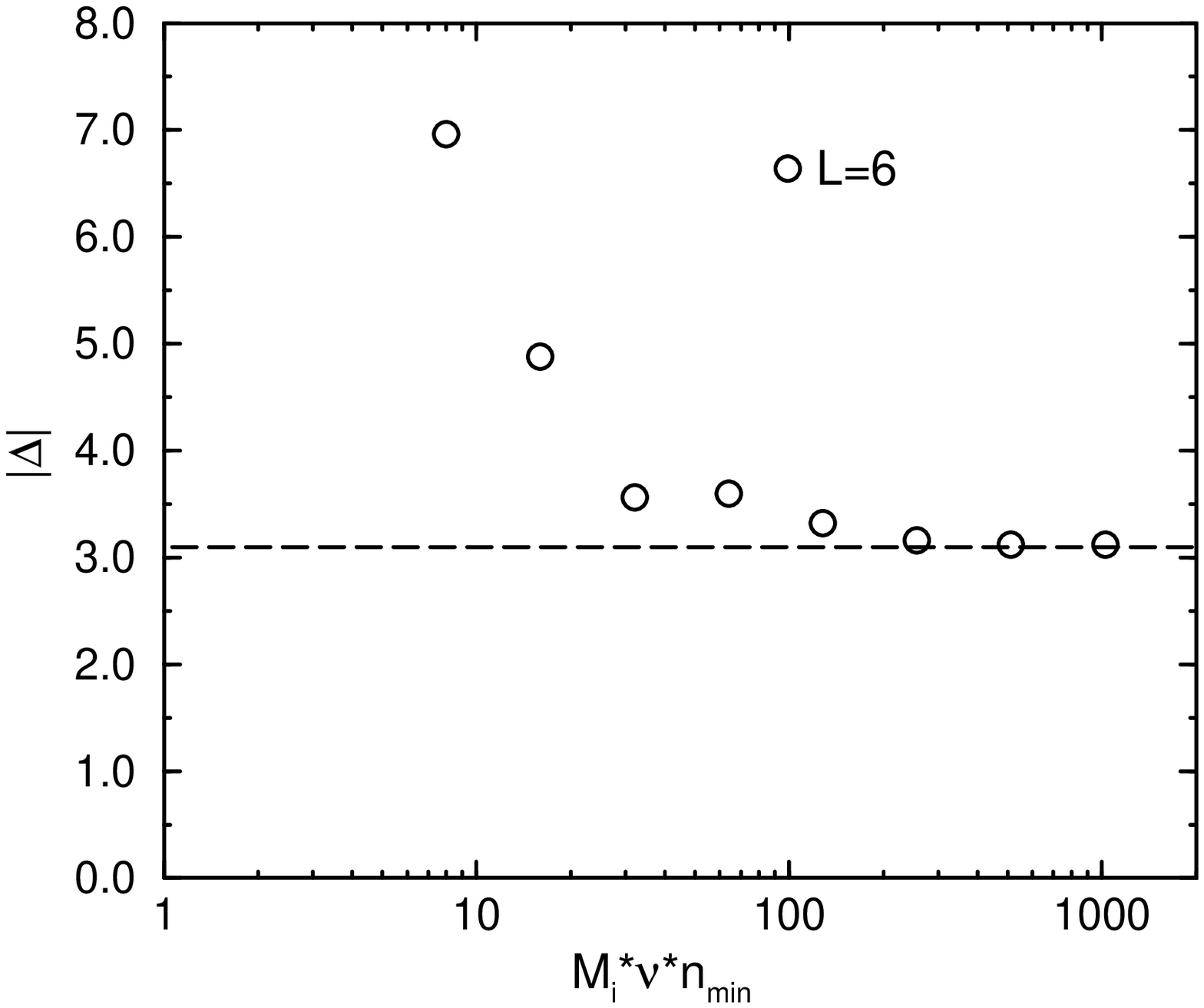}}
\end{center}
\caption{\captionC}
\label{fig_test_delta}
\end{figure}
}

\newcommand{\figD}{
\begin{figure}[ht]
\begin{center}
\myscalebox{\includegraphics{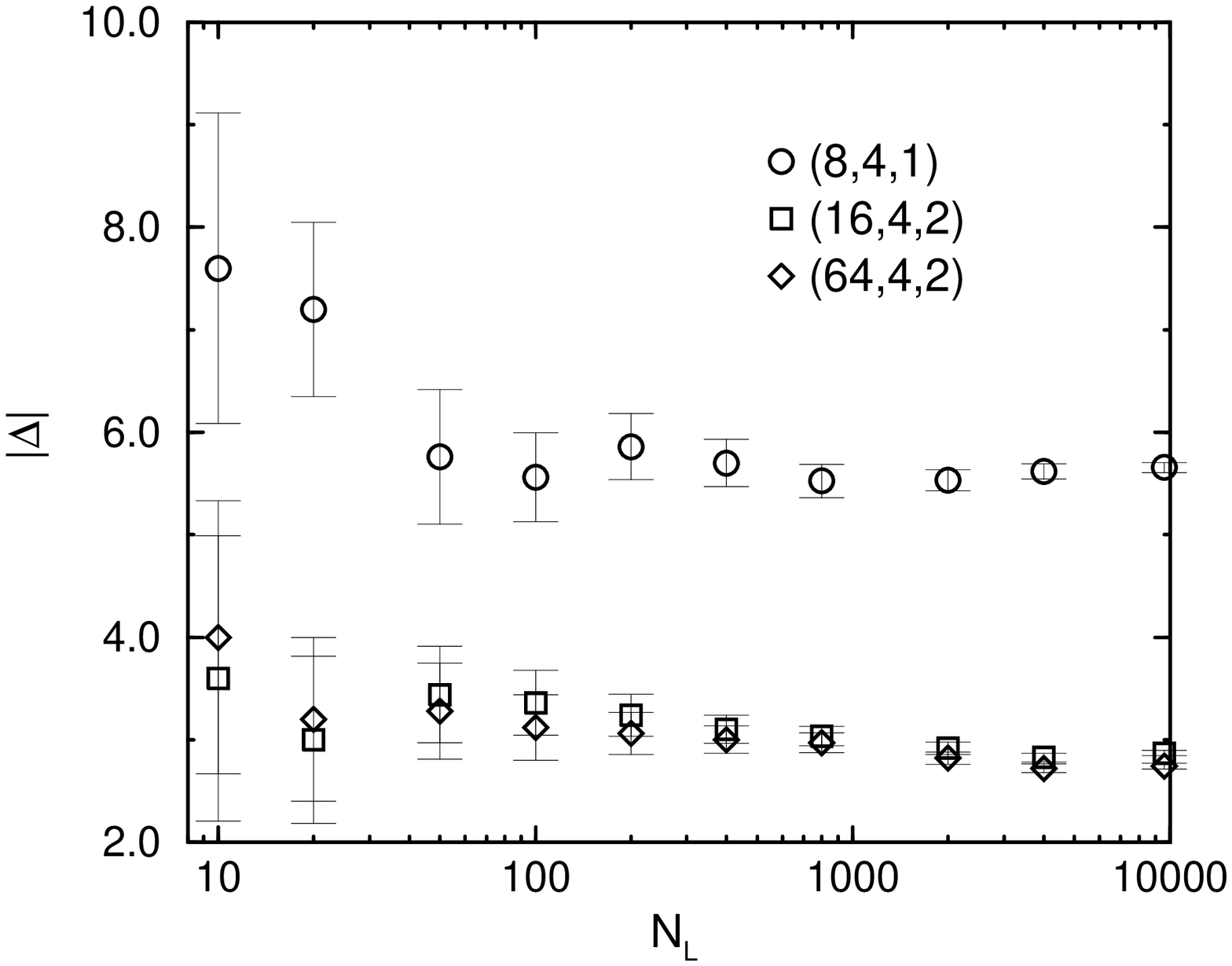}}
\end{center}
\caption{\captionD}
\label{fig_delta_e_n}
\end{figure}
}

\begin{document}
\title{Scaling of stiffness energy for 3d $\pm J$ Ising spin glasses}

\author{Alexander K. Hartmann\\
{\small  hartmann@tphys.uni-heidelberg.de}\\
{\small  Institut f\"ur theoretische Physik, Philosophenweg 19, }\\
{\small 69120 Heidelberg, Germany}\\
{\small Tel. +49-6221-549449, Fax. +49-6221-549331}}

\date{\today}
\maketitle
\begin{abstract}
Large numbers of ground states of 3d EA Ising spin glasses are calculated
for sizes up to $10^3$ using a combination 
of a genetic algorithm and Cluster-Exact
Approximation. A detailed analysis shows that true ground states are obtained.
The ground state stiffness (or domain wall) energy $\Delta$ is calculated. A
$|\Delta| \sim L^{\Theta_S}$ behavior with $\Theta_S=0.19(2)$ is found which
strongly indicates that the 3d model has an equilibrium 
spin-glass-paramagnet transition for non-zero $T_c$.

{\bf Keywords (PACS-codes)}: Spin glasses and other random models (75.10.Nr), 
Numerical simulation studies (75.40.Mg),
General mathematical systems (02.10.Jf). 
\end{abstract}


\paragraph*{Introduction}
The question whether three dimensional (3d) Ising spin glasses \cite{binder86}
 have
a non-zero transition temperature $T_c$ is still not answered beyond all
doubts. The best evidence  for ordering
below a finite $T_c$ was found recently \cite{kawashima96,marinari98} 
by extensive 
Monte Carlo Simulations. But the authors could not completely rule out
other scenarios.

In this work we address the question by calculating the stiffness or
domain wall energy $\Delta=E^a-E^p$ which is the difference between the ground
state energies $E^a, E^p$ for antiperiodic and periodic boundary conditions
\cite{bray84,mcmillan84}. This quantity was studied earlier only for very 
small system sizes using $T=0$ transfer matrix methods \cite{bray84,cieplak90}
 and Monte Carlo simulations \cite{mcmillan84}. The stiffness energy
shows a finite size dependence
\begin{equation}
|\Delta| \sim L^{\Theta_S}
\end{equation}
where $L$ is the linear system size.
A positive value of the stiffness exponent $\Theta_s$ indicates the
existence of a spin glass phase for non-zero temperature. 
Since the direct calculation of
ground states for 3d spin glasses is NP-hard there is no
polynomial algorithm available. In our work we calculate
ground states using a combination of Cluster-Exact Approximation
(CEA) \cite{alex2} and a genetic algorithm \cite{pal96,michal92}.
Similar calculations proved that the 2d spin glass, 
where exact ground states
can be calculated using a polynomial time algorithm, exhibits no ordering for 
$T>0$ \cite{kawashima97}.

We investigate systems of $N$ spins 
$\sigma_i = \pm 1$, described by the Hamiltonian
\begin{equation}
H \equiv - \sum_{\langle i,j\rangle} J_{ij} \sigma_i \sigma_j
\end{equation}
In this letter we
consider 3d cubic systems with periodic boundary conditions,
$N=L^3$ spins and the nearest neighbor interactions (bonds) take
independently $J_{ij} = \pm 1$ with equal probability. The antiperiodic 
boundary conditions for calculating $E^a$ are realized by inverting one
plane of bonds.

\paragraph*{Algorithm}
The algorithm for the calculation bases on a special genetic
algorithm \cite{pal96,michal92} and on Cluster-Exact Approximation  
\cite{alex2} which is  an sophisticated optimization method.
Now a short sketch of these algorithms is given, because later the
influence of different simulation parameters on the results is discussed.

The genetic algorithm starts with an initial population of $M_i$
randomly initialized spin configurations (= {\em individuals}),
which are linearly arranged in
a ring. Then $\nu \times M_i$ times two neighbors from the population
are taken (called {\em parents}) and two offsprings are created
using a triadic crossover: a mask is used which is a
third randomly chosen (usually distant) member of the population with
a fraction of $0.1$ of its spins reversed. In a first step the
offsprings are created as copies of the parents. Then those spins are selected,
 where the orientations of the
first parent and the mask agree \cite{pal95}. 
The values of these spins
are swapped between the two offsprings. Then a mutation with a rate of $p_m$
is applied to each offspring, i.e. a fraction $p_m$ of the
spins is reversed.

Next for both offsprings the energy is reduced by applying
CEA.
The method constructs iteratively and randomly 
a non-frustrated cluster of spins, whereas
spins with many unsatisfied bonds are more likely to be added to the
cluster. For 3d $\pm J$ spin glasses each cluster
contains typically 58 percent of all spins.
The  non-cluster spins act like local magnetic fields on the cluster spins.
For the spins of the cluster an energetic minimum state can be 
calculated in polynomial time
by using graph theoretical methods 
\cite{claibo,knoedel,swamy}: an equivalent network is constructed
\cite{picard1}, the maximum flow is calculated 
\cite{traeff,tarjan}\footnote{Implementation details: We used 
Tarjan's wave algorithm together
with the heuristic speed-ups of Tr\"aff. In the construction of 
the {\em level graph} we allowed not only edges $(v, w)$
with level($w$) = level($v$)+1, but also all edges $(v,t)$ where $t$
is the sink. For this measure, we observed an additional speed-up of
roughly factor 2 for the systems we calculated.} and the spins of the
cluster are set to their orientations leading to a minimum in energy. 
This minimization step
is performed $n_{\min}$ times for each offspring.

Afterwards each offspring is compared with one of its parents. The
pairs are chosen in the way that the sum of the phenotypic differences
between them is minimal. The phenotypic difference is defined here as the
number of spin positions where the two configurations differ. Each
parent is replaced if its energy is not lower (i.e. better) than the 
corresponding offspring.
After this whole step is done $\nu \times M_i$ times the population
is halved: From each pair of neighbors the configuration 
 which has the higher energy is eliminated. If not more than 4
individuals remain the process is stopped and the best individual
is taken as result of the calculation.

The following representation summarizes the algorithm. 
\vspace{0.0cm}

\newlength{\mpwidth}
\setlength{\mpwidth}{\textwidth}
\addtolength{\mpwidth}{-2cm}
\begin{center}

\begin{minipage}[b]{\mpwidth}
\newlength{\tablen}
\settowidth{\tablen}{xxx}
\newcommand{\tabspace}{\hspace*{\tablen}}
\begin{tabbing}
\tabspace \= \tabspace \= \tabspace \= \tabspace \= \tabspace \=
\tabspace \= \kill
{\bf algorithm} genetic CEA($\{J_{ij}\}$,
$M_i$, $\nu$, $p_m$, $n_{\min}$)\\
{\bf begin}\\
\> create $M_i$ configurations randomly\\
\> {\bf while} ($M_i > 4$) {\bf do}\\
\> {\bf begin}\\
\> \> {\bf for} $i=1$ {\bf to} $\nu \times M_i$ {\bf do}\\
\>\> {\bf begin}\\
\>\>\> select two neighbors \\
\>\>\> create two offsprings using triadic crossover\\
\>\>\> do mutations with rate $p_m$\\
\>\>\> {\bf for} both offsprings {\bf do}\\
\>\>\> {\bf begin}\\
\>\>\>\> {\bf for} $j=1$ {\bf to} $n_{\min}$ {\bf do}\\
\>\>\>\> {\bf begin}\\
\>\>\>\>\> construct unfrustrated cluster of spins\\
\>\>\>\>\> construct equivalent network\\
\>\>\>\>\> calculate maximum flow\\
\>\>\>\>\> construct minimum cut\\
\>\>\>\>\> set orientations of cluster spins\\
\>\>\>\> {\bf end}\\
\>\>\>\> {\bf if} offspring is not worse than related parent \\
\>\>\>\> {\bf then}\\
\>\>\>\>\> replace parent with offspring\\
\>\>\> {\bf end}\\
\>\> {\bf end}\\
\>\> half population; $M_i=M_i/2$\\
\> {\bf end}\\
\> {\bf return} one configuration with lowest energy\\
{\bf end}
\end{tabbing}

\end{minipage}
\end{center}
\vspace{0.5cm}

The whole algorithm is performed $n_R$ times and all configurations
which exhibit the lowest energy are stored, resulting in $n_g$ statistical
independent ground state configurations.

This algorithm was already applied to examine the ground state 
structure of 3d spin glasses \cite{alex_sg2}.

\paragraph*{Results}
For each system size we tried many different combinations
of the simulation
parameters $m_i, \nu, n_{min}, p_m$ for some sample systems. The final
parameters where determined in a way, that
by using four times the numerical effort no reduction in
energy was obtained. Here $p_m=0.2$ and $n_R=10$ were used
for all system sizes. 
Table \ref{tab_parameters} summarizes the parameters. Also the
typical computer time $\tau$ per ground state
computation on a 80 MHz PPC601 is given. 

Ground states were calculated for system sizes up to $L=10$ for $N_L$
independent realizations (see table \ref{tab_parameters})
of the random variables. For each realization
the ground states with periodic and antiperiodic boundary condition
were calculated. One can extract from the table that the $L=10$
systems alone required 1990 CPU-days. 
Using these parameters on average $n_g>8$
ground states were obtained for every system size $L$ using  $n_R=10$ runs per
realization.

In the second part of this paragraph a detailed analysis of the
influence of the simulation parameters is presented. 
But at first the results
for the stiffness energy are shown in Fig. \ref{fig_stiff_L}.


Also given is a fit $\Delta(L)\sim L^{\Theta_S}$ which results in 
$\Theta_S=0.19(2)$. Because of the large sample sizes the error bars are small
enough, so we can be pretty sure that $\Theta_S>0$. It means that the 3d
EA spin glass exhibits a non-zero transition temperature $T_c$. 
Since $\Delta$ is a small
difference of large values, we have to be sure that we calculate true
ground states in order to believe our results.


Fig. \ref{fig_test_e} shows the average energy per spin
for 90 test systems of
size $L=6$ as function of the pro\-duct of the
simulation parameters $M_i\times\nu 
\times n_{\min}$ which is proportional to the computer time since all
other parameters are kept fixed. 
The sets $(M_i,\nu,n_{\min})=(8,1,1)$,
$(8,2,1)$, $(8,4,1)$ and $M_i=8,16,32,64,128$ for $(\nu,n_{\min})=(4,2)$
were investigated.
The energy decreases with increasing numerical effort. For the sets with
$M_i>32$ the energy does not decrease further. We compared our results for 
 the test systems with exact ground states which were obtained
using a  Branch-and-Cut
program \cite{simone95,simone96}. For the two largest parameter sets
the genetic CEA algorithm found the true ground states for all 90 systems!
The same result was obtained for $L=4$ as well.\footnote{For $L>6$ the 
Branch-and-Cut needs to much computer time because of the exponential
time complexity.} So we can be sure that genetic CEA
and our method of choosing the parameters lead to true ground states or
at least to states very close to true ground states.


The choice of the simulation parameters has 
only a small influence on the the energy for large values of $M_i$. A 
more sensitive indicator is the stiffness energy as function of the simulation
parameters. This is shown for the same 90 test systems in Fig.  
\ref{fig_test_delta}. We also calculated the exact ground states for
the realizations with antiperiodic boundary conditions and found again
that genetic CEA produced 
the exact results for the parameter sets with $M_i>32$.

Since the parameters were tested always on a restricted number of systems we
can not absolutely be sure that genetic CEA always finds true ground states.
Since we are interested in the stiffness energy $\Delta$ we take a 
closer look at it. If
states very close to the true ground states are found, the resulting
 stiffness energy 
may be for some realizations smaller and for others 
higher than the correct result. 
So we expect that the effect should cancel out with increasing
sample size.


This is confirmed
 by Fig. \ref{fig_delta_e_n} where the stiffness energy
as function of the sample size $N_L$ is shown 
for three different parameter sets (8,4,1), (16,4,2) and (64,4,2).
The stiffness energies
found using the second set, where we are very close to the true ground states,
converges to the values found using the third set. The values found using the
first set do not converge. So even by calculating states very close to the
true ground states on gets very good estimates of the stiffness energy.

\paragraph*{Conclusion}
Results have been presented from calculations of 
a large number of ground states of 3d Ising spin glasses. They were
obtained
using a combination of Cluster-Exact Approximation
and a genetic algorithm. The finite size behavior of the
stiffness energy has been investigated. 
Is has been shown that states  which were obtained 
are true ground states or are at least very close
to them. Even if one calculates only almost true ground states the resulting 
value of the stiffness energy is very reliable. It has been found that 
the stiffness
energy increases with system size.  So  strong evidence has been obtained 
that the 3d Ising spin glass has a non-zero transition temperature $T_c$.

The only uncertainty arises from the fact that these calculations were
restricted to systems sizes $L\le 10$,  it means anyway that the number 
of spins is more than 200
times larger than for the systems examined before ($L=6$) 
\cite{bray84,mcmillan84,cieplak90}. Additionally in
contrast to former publications it is possible to
estimate the quality of the low energy states.

\paragraph*{Acknowledgements}

The author thanks H. Horner and G. Reinelt for manifold support.
He is grateful to M. J\"unger, M. Diehl and T. Christof who put
a Branch-and-Cut Program for the exact calculation of spin glass
ground states of small systems at his disposal.
He thanks R. K\"uhn for critical reading of the manuscript and for giving 
many helpful hints.
This work was engendered by interesting discussions with 
H. Rieger, A.P. Young, 
and N. Kawashima during the ``Algorithmic Techniques in Physics'' seminar held
at the {\em International Conference and Research Center for
Computer Science Schloss Dagstuhl} in Wadern/Germany.
The author took much benefit from discussions with H. Kinzelbach and S. Kobe.
He is also grateful to the {\em Paderborn Center for Parallel Computing}
 for the allocation of computer time. This work was supported
by the Graduiertenkolleg ``Modellierung und Wissenschaftliches Rechnen in 
Mathematik und Naturwissenschaften'' at the
{\em In\-ter\-diszi\-pli\-n\"a\-res Zentrum f\"ur Wissenschaftliches Rechnen}
 in Heidelberg.

\begin{table}[h]
\begin{center}
\begin{tabular}{cccccc}
\hline
$L$ & $M_i$ & $\nu$ & $n_{\min}$ & $\tau$ (sec) & $N_L$ \\ \hline
4 & 32 & 3 & 1 & 1 & 10000 \\
6 & 64 & 4 & 2 & 20 & 10000 \\
8 & 64 & 4 & 5 & 140 & 12469 \\
10 & 128 & 6 & 10 & 1920 & 4480 
\end{tabular}
\end{center}
\caption{Simulation parameters: $L$ = system size, $M_i$ = initial size of
population, $\nu$ = average number of offsprings per configuration, $n_{\min}$
= number of CEA minimization steps per offspring, $\tau$ = average computer
time per ground state on a 80MHz PPC601, $N_L$ = number of realizations
of the random variables.}
\label{tab_parameters}
\end{table}



\figA
\figB
\figC
\figD

\end{document}